\documentclass[lettersize,journal]{IEEEtran}
\usepackage{amsmath,amsfonts}
\usepackage{array}
\usepackage[caption=false]{subfig}
\usepackage{textcomp}
\usepackage{stfloats}
\usepackage{url}
\usepackage{verbatim}
\usepackage{graphicx}
\usepackage{cite}
\usepackage{pifont}
\usepackage[noend]{algpseudocode}

\usepackage{amsmath}
\usepackage{amssymb}
\usepackage{booktabs}
\usepackage{multirow}
\usepackage[table,xcdraw]{xcolor}

\usepackage[pagebackref,breaklinks,colorlinks]{hyperref}

\begin{document}

\title{AudioRepInceptionNeXt: A lightweight single-stream architecture for efficient audio recognition}

\author{Kin Wai Lau$^{1,2}$, Yasar Abbas Ur Rehman$^{2}$, Lai-Man Po$^{1}$ \\
City University of Hong Kong$^{1}$\\
TCL AI Lab$^{2}$\\
        % <-this % stops a space
\thanks{K.W. Lau is with the Department of Electrical Engineering, City University of Hong Kong, Hong Kong, and also with TCL AI Lab. Y.A.U. Rehman is with TCL AI Lab. (e-mail: kinwailau6-c@my.cityu.edu.hk, yasar.abbas@my.cityu.edu.hk)}% <-this % stops a space
\thanks{L.-M. Po is with the Department of Electrical Engineering, City University of Hong Kong, Hong Kong (email: eelmpo@cityu.edu.hk)}}

\markboth{Journal of \LaTeX\ Class Files,~Vol.~14, No.~8, Feb~2024}%
{Shell \MakeLowercase{\textit{et al.}}: A Sample Article Using IEEEtran.cls for IEEE Journals}


\maketitle
\newcommand{\cmark}{\ding{51}}
\newcommand{\xmark}{\ding{55}}

\begin{abstract}
Recent research has successfully adapted vision-based convolutional neural network (CNN) architectures for audio recognition tasks using Mel-Spectrograms. However, these CNNs have high computational costs and memory requirements, limiting their deployment on low-end edge devices. Motivated by the success of efficient vision models like InceptionNeXt and ConvNeXt, we propose AudioRepInceptionNeXt, a single-stream architecture. Its basic building block breaks down the parallel multi-branch depth-wise convolutions with descending scales of $k \times k$ kernels into a cascade of two multi-branch depth-wise convolutions. The first multi-branch consists of parallel multi-scale $1\times k$  depth-wise convolutional layers followed by a similar multi-branch employing parallel multi-scale  $k \times 1$  depth-wise convolutional layers. This reduces computational and memory footprint while separating time and frequency processing of Mel-Spectrograms. The large kernels capture global frequencies and long activities, while small kernels get local frequencies and short activities. We also reparameterize the multi-branch design during inference to further boost speed without losing accuracy. Experiments show that AudioRepInceptionNeXt reduces parameters and computations by 50\%+ and improves inference speed $1.28\times$ over state-of-the-art CNNs like the Slow-Fast while maintaining comparable accuracy. It also learns robustly across a variety of audio recognition tasks. Codes are available at \url{https://github.com/StevenLauHKHK/AudioRepInceptionNeXt}.
\end{abstract}


\section{Introduction}
\label{sec:intro}
% Write about the motivation first
Learning deep feature representations for audio understanding has been extensively studied over the past decade using a variety of deep neural network architectures like Convolutional Neural Networks (CNN) \cite{kazakos2021slow, hershey2017cnn, verbitskiy2022eranns, engel2017neural, allamy20211d}, Long-Short Term Memory (LSTM) \cite{lezhenin2019urban, duan2021short, utebayeva2020multi, das2020urban}, and the recent Transformer networks \cite{gong2021ast, gong2022ssast, chen2022hts}. These deep neural networks typically learn the mapping from an audio sample to its corresponding label, \textit{intermediate} feature representations \cite{saeed2021contrastive, niizumi2021byol, wang2021multi}, or augmented audio sample \cite{suh2019acoustic, park2019specaugment}. In practice, these deep neural networks for audio-understanding tasks have the flexibility to be trained by either using the raw audio samples \cite{baevski2020wav2vec, gong2021ast, gong2022ssast} or a 2D time-frequency spectrogram \cite{kazakos2021slow, ford2019deep, schmid2022efficient, gong2021psla, wang2022towards, kong2020panns, verbitskiy2022eranns, saeed2021contrastive, niizumi2021byol, wang2021multi}. Recent advances in deep neural networks have had a revolutionary impact on numerous audio understanding domains, including but not limited to predictive tasks like sound event classification \cite{mesaros2021sound}, the direction of voice prediction \cite{ahuja2020direction}, speech command recognition \cite{warden2018speech}, speaker identification \cite{nagrani2017voxceleb}, and generative tasks such as music generation \cite{chen2020music}. Although the Transformers-based networks such as wave2vec \cite{baevski2020wav2vec} and ViT \cite{gong2021ast} show promising results in these audio understanding tasks; deploying them in their na\"ive form on the edge devices would require allocating massive amounts of compute resources for the architecture besides the audio data. For example, the base model of  wave2vec 2.0 \cite{baevski2020wav2vec} requires over 89.78M (in millions) parameters compared to CNN models that only require 4.60M parameters  \cite{gao2022federated}. This limits the applicability of Transformers in realizing numerous recent applications of general-purpose audio understanding that require on-device computation and training, such as federated learning \cite{gaol2023match}. Except for speech recognition, we found that the CNN-based deep neural networks are still prevalent for audio understanding tasks, such as audio event recognition and music classification, while maintaining similar performance compared to Transformers, and suitable for deployment and running on edge devices  \cite{lau2023audioinceptionnext}. 

\begin{figure*}[t]
\centering
\includegraphics[width=0.7\linewidth]{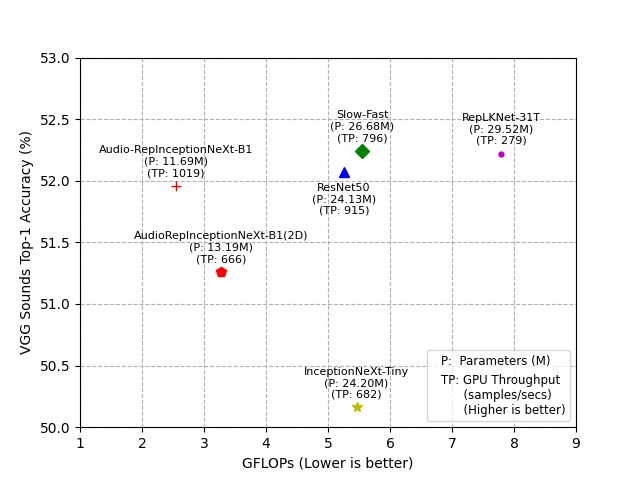}
\caption{Comparison of the Top-1 accuracy and GFLOPs on VGG Sounds. Different markers represent different baseline backbone architectures.}
\label{fig:flops-accuracy-classification}
\end{figure*}

The Slow-Fast \cite{kazakos2021slow} is a recent framework focusing on the CNN architecture design for audio understanding, which proposed a two-stream pathway CNN, that has obtained better performance with lower parameter count than the Transformers on EPIC-SOUND datasets \cite{lau2023audioinceptionnext}. Following the success of separable kernels in the recent work on audio recognition \cite{xiao2020audiovisual}, the Slow-Fast model proposed to use $1 \times k$ and $k \times 1$ kernels to capture the frequency and temporal feature independently considering  
 non-homogeneous statistics of the audio-spectrogram. Later work \cite{wang2022towards} extends the study of the Slow-Fast model with self-supervised contrastive learning and discovered that such architectures provide better feature generalization on a diverse set of audio understanding tasks. Although performing remarkably well on a variety of audio understanding tasks, we found that these CNN models still incur high computational costs and memory footprints that potentially limit their applicability on edge devices for a variety of audio understanding applications. As an example, the Slow-Fast model incurs 26.68M parameters, which is 1.10$\times$ higher than the conventional ResNet-50 model (with 24.13M parameters) (See Figure \ref{fig:flops-accuracy-classification}).  

To enable general-purpose audio understanding on edge devices without incurring high computational and memory footprints, we focus here on a parallel rather unexplored area of redesigning and reparameterization of CNN architectures. Our motivation for redesigning and investigating the CNN-based architectures for general-purpose audio understanding is due to the following: (1) Although CNN architectures incur lower computational and memory costs, their performance heavily depends on the network architecture and design. In practice, direct deployment of the trained model on the low-end edge devices for inference might result in slow inference speed and increased memory footprints. (2) The Siamese networks, such as the Slow-Fast \cite{kazakos2021slow}, incur a higher computational and memory footprint than single-stream networks like ResNet50, InceptionNeXt-Tiny and our proposed network (see. Figure \ref{fig:flops-accuracy-classification}). (3) The single-stream multi-branch CNN \cite{lau2023audioinceptionnext} networks with reduced parameters and theoretical GLOPS show low throughput (audio frames/seconds) due to the high memory access costs \cite{ding2022scaling, ding2021repvgg, ma2018shufflenet},  and inefficient configuration of small operators in the multi-branch design \cite{ma2018shufflenet}.

\begin{figure*}[t!]
    \centering
    \subfloat[Slow-Fast Model vs. AudioRepInceptionNeXt]{\includegraphics[width=0.75\linewidth, height=0.45\linewidth]{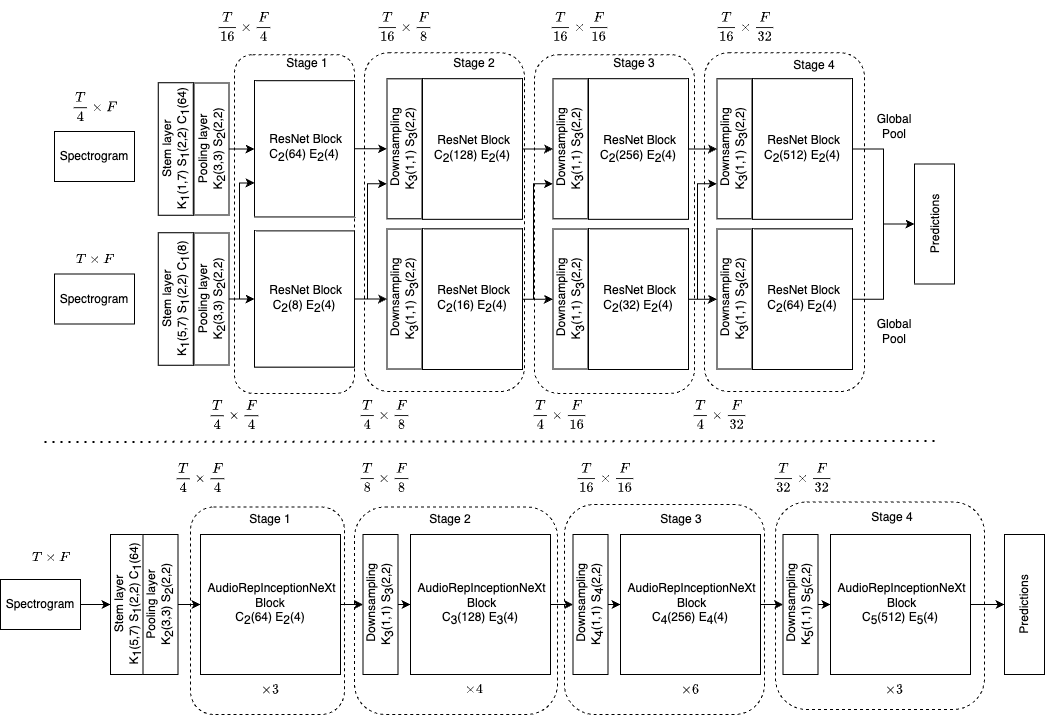}
    \label{fig:AudioConvNeXt-arch-a}} \\
    \subfloat[AudioRepInceptionNeXt Block (Left); AudioRepInceptionNeXt (2D) Block (Right)]{\includegraphics[width=0.6\linewidth, height=0.4\linewidth]{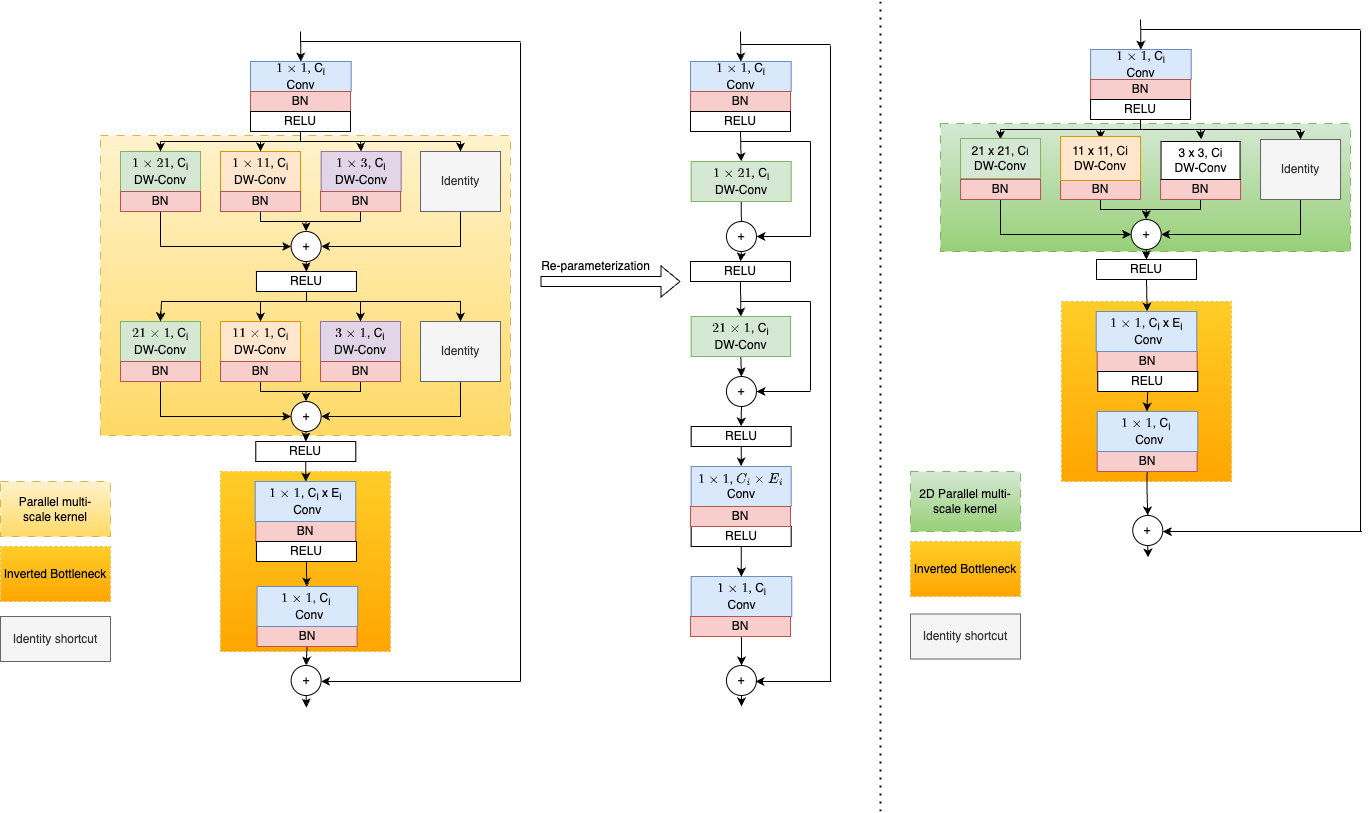}
    \label{fig:AudioConvNeXt-block-b}}
    \caption{(a) Architecture of the Slow-Fast Model (Upper) and AudioRepInceptionNeXt (Bottom); (b) AudioRepInceptionNeXt Block during the training (Left) and Structural Re-parameterization of AudioRepInceptionNeXt Block after the training (Right); AudioRepInceptionNeXt (2D) (Right side of the dotted line) DW-Conv represents the depth-wise convolution and other notations can be found in Section \ref{sec:methodology}.}
\label{fig:AudioConvNeXt-arch}
\end{figure*}

To address these issues, we focus on the design of a very deep and parameter-efficient CNN architecture for general-purpose audio recognition tasks that can be easily deployed on edge devices. Unlike the Slow-Fast model proposed recently \cite{kazakos2021slow}, we rethink the design of CNN for audio recognition and propose an efficient single-stream CNN architecture called AudioRepInceptionNeXt by employing parallel multi-scale separable convolutional kernels (see. Figure \ref{fig:AudioConvNeXt-block-b}).

The proposed model incurs lower computational and memory footprints while maintaining similar performance as the state-of-the-art CNN models. Additionally, the parallel multi-scale convolutional kernels in the proposed model can be rescaled to a single-scale convolutional branch during the inference time that not only further reduces the computational and memory footprints but also enhances the network throughput by a significant margin while maintaining the same performance as the original design.  In this way, the model can capture the local and global temporal-frequency information via the multi-scale kernel designs during the training while eliminating the side-effect of multi-scale kernel design (i.e., slow inference speed and high memory access costs) during the inference time. Such design allows the simultaneous use of very large-scale kernels, e.g., $21 \times 21$, and small-scale kernels, e.g., $3 \times 3$ \cite{ding2022scaling, guo2022visual, liu2022convnet} in the parallel multi-scale branch of AudioRepInceptionNeXt during training. This design also allows our model to capture the global frequency semantic information and long-duration activities, and local details of frequency information and short-duration activities. We found that the proposed design takes few parameters (26.68M vs. 11.69M), lower computational complexity (5.55 GFLOPs vs. 2.55 GFLOPs), and higher inference speed (796 samples/sec vs. 1019 samples/sec) compared to the two-stream Slow-Fast model while achieving similar performance with a marginal difference of 0.28\%  in accuracy (see Figure \ref{fig:flops-accuracy-classification}).  \\  

Our contributions can be summarized as follow:
\begin{enumerate}
    \item We address the computational inefficiency issues in the multi-stream Slow-Fast model. We show that the proposed single-stream multi-scale separable kernel architecture, AudioRepInceptionNeXt, effectively reduces the number of parameters and computational complexity incurred by multi-stream network architecture, without any performance degradation.
    \item We employ reparameterization techniques \cite{ding2022scaling, ding2021repvgg, ding2021diverse} to further eliminate the side effect of multi-scale kernel design (i.e., slow inference speed and high memory access costs) by converting it into a single separable kernel during the inference, without any performance drop.
    \item We validate the effectiveness of the proposed AudioRepInceptionNeXt on various audio classification tasks, including sound event classification, speech command classification, and music instrument classification. We demonstrate that AudioRepInceptionNeXt can achieve comparable or superior results as the state-of-the-art CNN-based models on the variety of downstream tasks \cite{chen2020vggsound} while saving half of the GFLOPs and memory footprint.
\end{enumerate}

The rest of this paper is organized as follows. In Section \ref{sec:related-work}, we introduce literature on the state-of-the-art single-stream and multi-stream CNN-based network for audio recognition. In addition, we introduce the motivation for using model reparameterization in this paper. Section \ref{sec:methodology} presents our proposed AudioRepInceptionNeXt network, followed by the experiment results in Section \ref{sec:experiment}. Section \ref{sec:Ablation Studies} provides ablation studies with the multi-branch design and the usage of large kernels. Finally, we conclude our work with a conclusion and future work in Section \ref{sec:conclusion}.

\section{Related Work}
\label{sec:related-work}
\subsection{Single-Stream Architecture to Multi-Stream Architecture}
Single-stream convolution neural network (CNN) is widely used in audio recognition tasks including sound event classification, speech recognition, and music classification \cite{salamon2017deep, hershey2017cnn, aytar2016soundnet, martin2022low, schneider2019wav2vec}. There are two widely used approaches in single-stream CNN-based audio recognition systems. The first approach treats the raw audio waveform as input and utilizes a single-stream CNN to extract the feature for classification \cite{schneider2019wav2vec, baevski2020wav2vec, li2019feature, allamy20211d}. Wav2vec \cite{schneider2019wav2vec} is a representation work in this regard that takes the 1D audio signal as input and trains a 1D CNN in an unsupervised contrastive manner to learn the \textit{intermediate} audio representation for speech recognition.
The second approach first preprocesses the 1D audio signal into a 2D time-frequency Mel-spectrogram followed by feeding it to the 2D CNN \cite{schmid2023efficient, verbitskiy2022eranns, gong2021psla, jansen2018unsupervised}.

On the other hand, multi-stream architectures simultaneously work with either single or multimodal data such as raw audio and spectral features of audio \cite{kong2020panns, li2019multi, kazakos2021slow}. Slow-Fast model \cite{kazakos2021slow} is a representative work that uses two-stream network architecture to capture both frequency and fine-grained temporal information from two different resolutions of the Log-Mel-Spectrogram independently. Their results demonstrated that the performance of the two-stream network outperforms the state-of-the-art single-stream network. 

Although the above-mentioned single-stream and two-stream networks demonstrate promising results in audio recognition, they incur high computational and memory footprints during training and inference, which hinder their deployment on low-resource edge devices, such as smartphones. Therefore, it is imperative to explore a lightweight and parameter-efficient single-stream network that can achieve comparable results with the single-stream and two-stream networks while achieving low computational and memory footprints. In this work, we propose a parallel multi-scale convolutional kernel block to enrich the temporal and frequency feature representation during the training. This results in a single-stream CNN-based network architecture with lower parameters and computational footprints. The proposed design further enables the use of model reparameterization by combining multiple kernels into a single kernel during inference without any loss of performance.

\subsection{Model Re-parameterization}
Model Reparameterization \cite{ding2022scaling, ding2021repvgg, ding2021diverse} approaches simplify complex multi-branch network architecture into a single-branch network structure during the model inference stage, without sacrificing performance. These approaches enable the use of the complex network architecture for learning efficient feature representations during the training stage and a simplified and parameters-efficient network during the inference stage. For instance, RepVGG \cite{ding2021repvgg} proposed extra $1 \times 1$ convolutional layers parallel to the $3 \times 3$ convolution layers in the original VGG network to learn richer feature representations during the training. After the training, the additional $1\times 1$ layers are merged with the $3\times3$ layers via linear transformations of convolution, resulting in a VGG-like model with no extra parameters or computational cost during inference. RepVGG has been shown to outperform the original VGG model without adding any inference-time cost. Similarly, the Diverse Branch Block (DBB) method proposed in \cite{ding2021diverse} enhances the representation capacity of a single convolution by combining multiple branches of varying complexities to enrich the feature representation. The DBB block includes a sequence of convolutions, multi-scale convolutions, and average pooling during training, which can be converted into a single convolutional layer using the linear transformation properties of convolution. Inspired by the success of RepVGG and DBB, RepLKNet \cite{ding2022scaling} employed a similar technique by using a large kernel and a small kernel in parallel during training. After training, the small kernel is absorbed into the large kernel, enabling the large kernel to capture both global and local information resulting in improved the model's performance. 

Our motivation for model reparameterization differs from these previous works in one important way. \textit{We have discovered that the inference speed of the multi-branch architecture designs without reparameterization does not correlate closely with the number of model parameters and GLOPs} (see. Table \ref{tab:baseline-comparsion}). For example, the multi-branch architecture designs before reparameterization, such as InceptionNeXt\cite{yu2023inceptionnext} and the proposed AudioRepInceptionNeXt,
is slower compared to other models like the multi-stream Slow-Fast model, despite having lower parameters and GFLOPs. We attribute this discrepancy in speed to the increased memory access and synchronization time required by the multi-branch design \cite{ma2018shufflenet}. Therefore, the reparameterization approach adopted in our work focuses on addressing the issue of optimizing the memory access and synchronization time of multi-branch designs. By optimizing memory access and synchronization time in this way, the reduction in the number of model parameters and GLOPs becomes highly correlated with the increase in inference speed in multi-branch CNN design.

\section{Methodology}
\label{sec:methodology}

\begin{table*}[t]
    \caption{Detail settings of AudioRepInceptionNeXt.}
    \centering
    \resizebox{1.5\columnwidth}{!}{
    \begin{tabular}{l|c|c|c|c}
        \textbf{} & \textbf{Output Size} & 
        \textbf{Layer name} &
        \textbf{AudioRepInceptionNeXt-B0} &
        \textbf{AudioRepInceptionNeXt-B1} \\ 
        \hline
        \multirow{4}{*}{Stage 1} & \multirow{4}{*}{$\frac{T}{4} \times \frac{F}{4}$} & stem  & \multicolumn{2}{c}{$S_1=(2,2); K_1=(5,7)$} \\
        \cline{3-5}
        & & Max pooling layer & \multicolumn{2}{c}{$S_2=(2,2); K_2=(3,3)$}\\
        \cline{3-5}
        & & Embed. Dim & $C_1=32$ & $C_1=64$ \\
        \cline{3-5}
        & & Convolution Encoder & 
        \begin{tabular}{@{}c@{}}$C_2=32$ \\ $L_2=3$ \\ $E_2=4$\end{tabular} &
        \begin{tabular}{@{}c@{}}$C_2=64$ \\ $L_2=3$ \\ $E_2=4$\end{tabular} 
        \\
        \hline
        \multirow{4}{*}{Stage 2} & \multirow{4}{*}{$\frac{T}{8} \times \frac{F}{8}$} & Downsampling  & \multicolumn{2}{c}{$S_3=(2,2); K_3=(1,1)$} \\
        \cline{3-5}
        & & Convolution Encoder & 
        \begin{tabular}{@{}c@{}}$C_3=64$ \\ $L_3=4$ \\ $E_3=4$\end{tabular} &
        \begin{tabular}{@{}c@{}}$C_3=128$ \\ $L_3=4$ \\ $E_3=4$\end{tabular}
        \\
        \hline
        \multirow{4}{*}{Stage 3} & \multirow{4}{*}{$\frac{T}{16} \times \frac{F}{16}$} & Downsampling & \multicolumn{2}{c}{$S_4=2; K_4=(1,1)$} \\
        \cline{3-5}
        & & Convolution Encoder & 
        \begin{tabular}{@{}c@{}}$C_4=128$ \\ $L_4=6$ \\ $E_4=4$\end{tabular} &
        \begin{tabular}{@{}c@{}}$C_4=256$ \\ $L_4=6$ \\ $E_4=4$\end{tabular}
        \\
        \hline
        \multirow{4}{*}{Stage 4} & \multirow{4}{*}{$\frac{T}{32} \times \frac{F}{32}$} & Downsampling & \multicolumn{2}{c}{$S_1=(2,2); K_1=(1,1)$} \\
        \cline{3-5}
        & & Convolution Encoder & 
        \begin{tabular}{@{}c@{}}$C_5=256$ \\ $L_5=3$ \\ $E_5=4$\end{tabular} &
        \begin{tabular}{@{}c@{}}$C_5=512$ \\ $L_5=3$ \\ $E_5=4$\end{tabular} 
        \\
        \hline
        \end{tabular}
    }
    \label{tab:model-instantiation}
\end{table*}

In this section, we first describe the macro architecture design of the proposed AudioRepInceptionNeXt, followed by the micro block design in detail. We then provide a complexity analysis of the AudioRepInceptionNeXt Block.

\subsection{Model Architecture}
In this work, we use a typical hierarchical architecture design
% ResNet50\cite{he2016deep} 
as depicted in Fig. \ref{fig:AudioConvNeXt-arch} and the details are listed in Table \ref{tab:model-instantiation}. The hyperparameters of the model are listed as follows:
\begin{itemize}
\item {${S_i}$: the stride used in the convolutional layer in the input stem and in the downsampling layers in stage $i$;}
% \vspace{-0.2cm}
\item {${K_i}$: the kernel size of the convolutional layer in the input stem and in the downsampling in stage $i$;}
% \vspace{-0.2cm}
\item {${C_i}$: the number of output channels of stage $i$;}
% \vspace{-0.2cm}
\item {${E_i}$: the channel expansion ratio of inverted bottleneck in stage $i$;}
\item {${L_i}$: the number of blocks in stage $i$;}
% \vspace{-0.2cm}
\end{itemize}

\subsubsection{Model Input} As depicted in Figure. \ref{fig:AudioConvNeXt-arch-a}, AudioRepInceptionNeXt accepts the 2D audio mel-spectrogram as input with resolution $T \times F$. Unlike images, the width and height of the Mel-spectrogram represent distinct information, in which $T$ and $F$ axes correspond to the time and frequency bin, respectively. The time axis is generally longer than the frequency axis.

\subsubsection{Marco Design} Adhering to the design of ResNet50\cite{he2016deep}, our model comprises an input stem layer and four subsequent AudioRepInceptionNeXt stages. The first stem layer consists of a $5 \times 7$ convolutional layer with a stride of 2 on both the time and frequency axis and output feature maps with 64 channels. It is followed by a $3 \times 3$ Max-pooling layer with stride 2. The spatial resolution of the output feature maps after these two layers is 4 times lower compared to the spatial resolution of input to the network. Except for stage 1, each stage starts with the downsampling convolutional layer having  $1 \times 1$ kernel and a stride of 2. The downsampling layer is subsequently followed by an AudioRepInceptionNeXt block that contains a $1 \times 1$ convolution layer and a parallel separable kernel with kernel sizes of 21, 11, and 3. Similar to ConvNeXt \cite{liu2022convnet}, we adopted an inverted bottleneck design in each of the AudioRepInceptionNeXt blocks, where the width of the $1 \times 1$ MLP layer (expansion layer) is four times wider (along the channel dimensions) than the width of the input (along the channel dimension), as shown in Fig.\ref{fig:AudioConvNeXt-block-b} (left).

\subsection{AudioRepInceptionNeXt Block}
In this section, we describe the main components of the AudioRepInceptionNeXt block.

 \subsubsection{Parallel multi-scale kernel} As mentioned in Section \ref{sec:intro}, the key property of the Slow-Fast model is utilizing two streams of the CNN network in parallel to capture the temporal and frequency information at different scales. In contrast, we adopted the multi-branch design, motivated by the visual CNN-based InceptionNet \cite{szegedy2016rethinking}, which allows us to capture the temporal and frequency information at multiple scales with a single stream network. However, unlike the InceptionNet, we aim to use large-size kernels in the multi-branch layers, e.g., $21\times21$ and $11 \times 11$, in conjunction with small-size kernels, e.g., $3\times3$ and $1\times1$.  The large kernel (i.e., $21 \times 21$ and $11 \times 11$) captures the global-frequency semantic information and long-duration activities, while the small kernel (i.e., $3 \times 3$) captures the local-frequency information and short-duration activities. Finally, all the feature maps are added and passed to the $1 \times 1$ layers for channel-wise information exchange. However, na\"{i}vely using the large kernels may incur high computational costs. To tackle this issue, we use depthwise separable convolutional kernel design as explained in the following subsection.   
 
 \subsubsection{Depthwise Separable Kernel} As the parallel large kernel convolutional layers incur high computational costs in terms of the number of network parameters, we adopt the separable kernel design with depth-wise convolution (DW-Conv) following \cite{kazakos2021slow, xiao2020audiovisual}.   It should be noted that our design of depthwise separable kernels refers to the \textit{decomposition} of the 2D depthwise kernel and not to the conventional 2D depthwise separable kernels \cite{howard2017mobilenets} (that employ 2D depthwise convolution following the $1\times1$ convolution). 
 We decompose the 2D depthwise convolutional kernels, $k \times k$,  into $1 \times k$  and $k  \times 1$ as shown in Fig.\ref{fig:AudioConvNeXt-block-b} (left). However, instead of using a cascaded $1 \times k$ and $k \times 1$ architecture design, we first aggregate the multi-scale temporal features obtained by applying $1\times k$, followed by aggregating the multi-scale frequency features obtained by applying the $k\times1$ kernel. This design helps in obtaining two separable kernels  $1\times k$, and $k\times1$ after applying the reparameterization technique resulting in faster inference speed. In addition to reducing the memory footprint and computational time, previous studies \cite{kazakos2021slow, xiao2020audiovisual} have demonstrated that the use of such types of separable kernels allows the model to extract temporal and frequency information independently, leading to improvements in audio classification tasks. The reason is that the statistics of the spectrogram are not homogeneous, unlike natural images. We conducted a similar experiment within our network to verify the advantages of utilizing separable kernels, as discussed in Section \ref{sec:performance-on-pretrain}.
 
 \subsubsection{Inverted Bottleneck} 
 In the conventional design of the inverted bottleneck  \cite{sandler2018mobilenetv2},\cite{xie2017aggregated},  the number of channels in the hidden layer was four times the input channels. However, with large kernels e.g., $21$, and $11$, this design incurs increased computational cost.  Unlike these methods, we place the parallel multi-scale depthwise separable kernel at the top before applying the $1 \times 1$ expansion layer with expansion ratio ${E_i}$ for channel-wise information exchange following the ConvNeXt design. 

 \subsubsection{Identity shortcut}  Shortcut connections make the model an implicit ensemble of numerous shallower models \cite{veit2016residual, ding2022scaling}, such that the model can benefit from the different receptive fields. We demonstrate that shortcuts can improve the performance of AudioRepInceptionNeXt by 0.67\% in VGG-Sound audio event classification \cite{chen2020vggsound} as shown in section \ref{sec:Ablation Studies}.
 
\begin{figure*}[t]
\centering
\includegraphics[width=0.75\linewidth]{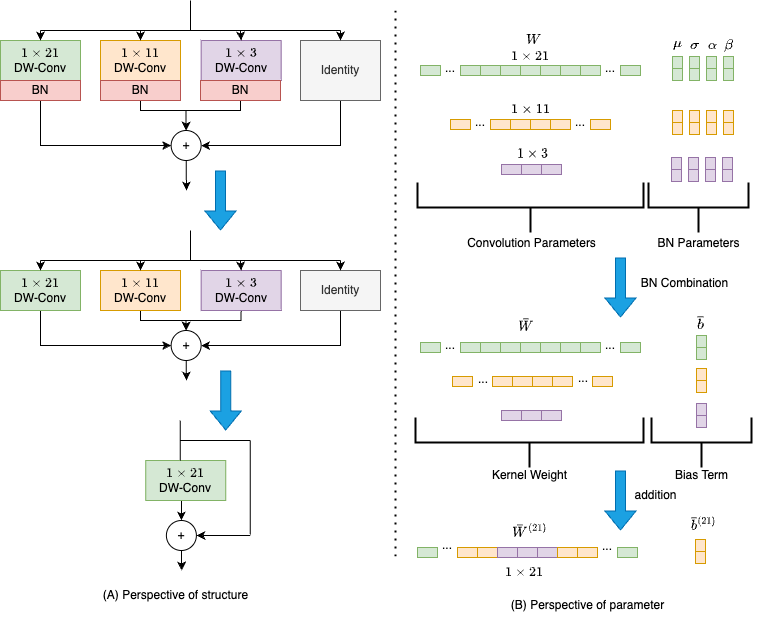}
\caption{Re-parameterization of the horizontal multi-scale kernel in the AudioRepInceptionNeXt Block. Here we assume all the layers have the same number of input channels, output channels, and stride size.}
\label{fig:Reparam}
\end{figure*}

\subsection{Reparameterization for Inference time model}
\label{subsec:rep-inference-model}
We employ the AudioRepInceptionNeXt depicted in Figure \ref{fig:Reparam}(a) during the training stage to learn feature representations from audio signals. During the inference stage, we first apply the reparameterization technique to convert the multi-branch AudioRepInceptionNeXt blocks into a single-branch reparametrized block as shown in Figure \ref{fig:Reparam}(b).

In this subsection, we describe how to convert the trained multi-branch kernels with varying scales into a single kernel (i.e., $1 \times k$ and $k \times 1$) for inference as shown in figure \ref{fig:Reparam}(b). Here we use the horizontal $1 \times k$ kernels as an example. A similar operation can be applied to the vertical $k \times 1$ kernel. 
One can see from Figure \ref{fig:Reparam}(b) we use three horizontal kernels of size $1\times21$, $1\times11$, and $1\times3$. Specifically, we use $W^{(21)}\in\mathbb{R}^{C_{in}\times C_{out}\times 1 \times 21}$ to denote the kernel of a $1 \times 21$ convolution layer with $C_{in}$ input channels and $C_{out}$ output channels, $W^{(11)}\in\mathbb{R}^{C_{in}\times C_{out}\times 1 \times 11}$ and $W^{(3)}\in\mathbb{R}^{C_{in}\times C_{out}\times 1 \times 3}$ for kernel $1 \times 11$ and $1 \times 3$, respectively. We use $\mu^{(i)}$, $\sigma^{(i)}$, $\alpha^{(i)}$, and $\beta^{(i)}$, where $i \in {3, 11, 21}$, as the mean, standard deviation, learned scaling factor and bias of the batch normalization (BN) layer following the $1 \times 21$, $1 \times 11$ and $1 \times 3$ convolution layers. Let $F^{(1)}\in\mathbb{R}^{N \times C_{in} \times H_{1} \times W_{1}}$ and $F^{(2)}\in\mathbb{R}^{N \times C_{out} \times H_{2} \times W_{2}}$ be the input and output features of the multi-scale horizontal convolution layers, respectively. $N$, $H$ and $W$ represent the batch size, height and width of the feature map, respectively. We assume $C_{in} = C_{out}$, $H_{1} = H_{2}$, and $W_{1} = W_{2}$ for simplifying the calculation. Before the re-parameterization, the output of the multi-branch horizontal convolution layers can be obtained by using the following equation.
% \begin{equation}
%      F^{(2)}= \alpha^{21} \times \frac{F^{(1)} * W^{(21)} - \mu^{(21)}}{\sqrt{\sigma^{2(21)}+\epsilon}} + \beta^{(21)} 
% \end{equation}

\begin{equation}
\begin{array}{l}
F^{(2)} = BN \{F^{(1)} * W^{(21)}, ~\mu^{(21)}, ~\sigma^{(21)}, ~\alpha^{(21)}, ~\beta^{(21)} \} \\ 
~~~\ + BN \{F^{(1)} * W^{(11)}, ~\mu^{(11)}, ~\sigma^{(11)}, ~\alpha^{(11)}, ~\beta^{(11)}\} \\ 
~~~\ + BN\{F^{(1)} * W^{(3)}, ~\mu^{(3)}, ~\sigma^{(3)}, ~\alpha^{(3)}, ~\beta^{(3)}\}, \\
    \label{eq:before-reparam}
\end{array}
\end{equation}
where 
\begin{equation}
\label{eq:bn}
BN(F,\mu,\sigma,\alpha,\beta)_{:,j,:,:} = (F_{:,j,:,:} - \mu_{j})\frac{\alpha_{j}}{\sigma_{j}} + \beta_{j}.
\end{equation}
Note that the identity branch is ignored in the equation \ref{eq:before-reparam} for simplifying the calculation. In equation \ref{eq:before-reparam}, the $*$ represents the convolution operation. In equation \ref{eq:bn}, $BN(.)$ and $j$ represent the batch normalization function and output channel index, respectively. $F_{:,j,:,:}$ represents the $j^{th}$ feature map output by the layer preceding the batch normalization layer.  Note that the sub-indices of the $BN(.)$ function and F follow the following order: batch size, output channel, height of the feature map, and width of the feature map. We first convert every BN layer and its corresponding horizontal convolution layer into a single convolution layer with a bias term. Let $\bar{W}^{(i)}$ and $\bar{b}^{(i)}$ be the kernel and the bias term after the combination, respectively. The kernel weight and bias can be obtained via the following equation.
\begin{equation}
\bar{W}^{(i)}_{j,:,:,:} = \frac{\alpha_{j}}{\sigma_{j}}W_{j,:,:,:},
\end{equation}
\begin{equation}
\bar{b}^{(i)}_{j} = -\frac{\mu_{j}\alpha_{j}}{\sigma_{j}} + \beta_{j}.
\end{equation}
Note that the sub-indices of the kernel weight $\bar{W}^{(i)}_{j,:,:,:}$ follow the following order: output channel, input channel, height of the kernel, and width of the kernel. After the combination, the output of each branch can be obtained by the following equation.
\begin{equation}
\begin{array}{l}
BN(F^{(1)}*W^{(i)}, \mu^{(i)}, \sigma^{(i)}, \alpha^{(i)}, \beta^{(i)})_{:,j,:,:} \\ = (F^{(1)}*\bar{W}^{(i)})_{:,j,:,:} 
+ \bar{b}^{(i)}_{j}.
\end{array}
\end{equation}
After the combination of the BN layer and its corresponding convolution layer, we can obtain three convolutional layers with horizontal kernel and three bias terms. Then we obtain the final $1 \times 21$ kernel by adding the $1 \times 11$ and $1 \times 3$ onto the central point of the $1 \times 21$ kernel and the final bias term by adding three bias terms together. The final kernel weight and bias can be obtained by the following equation.
\begin{equation}
\bar{W}^{(21)}_{j,:,:,:} = \sum_{n=1}^{i} \frac{\alpha^{n}_{j}}{\sigma^{n}_{j}}W^{n}_{j,:,:,:},
\end{equation}
\begin{equation}
\bar{b}^{(21)}_{j} = \sum_{n=1}^{i} -\frac{\mu^{n}_{j}\alpha^{n}_{j}}{\sigma^{n}_{j}} + \beta^{n}_{j}.
\end{equation}
Before adding both the small kernels to the large kernel, we apply zero padding to both small kernels such that they have the same kernel size as $1 \times 21$. Note that such transformation requires the $1 \times 11$ and $1 \times 3$ layer to have the same stride.

\subsection{Comparison to Multi-stream Slow-Fast model} 
Compared to the current state-of-the-art CNN-based Slow-Fast model\cite{kazakos2021slow}, our new proposed network uses a single-stream architecture instead of two-stream architecture. Our $21 \times 1$ and $1 \times 21$ large separable kernel can focus on the global-frequency semantic information and long-duration activities inherent in the similar functionality of the Slow model. Meanwhile, the $3 \times 1$ and $1 \times 3$ separable kernels act as a Fast model that captures the local-frequency semantic information and short-duration activities. The major benefit of using the single-stream network is that the computational complexity and memory footprint can be reduced by 54\% and 56\%, respectively, compared to the two-stream slow-fast model. In contrast, our network can achieve comparable performance as the slow-fast model. The following section will provide the complexity analysis of the new proposed architecture.

\section{Experiment}
\label{sec:experiment}
\subsection{Pretraining Dataset}
\textbf{VGG-Sound.} VGG-Sound \cite{chen2020vggsound} is a large-scale audio dataset extracted from YouTube videos. It contains more than 200K audio clips each with 10 seconds duration sampled at 16KHz. There are a total of 309 classes which include the sound emitted from objects, human actions, and interactions. 

\subsection{Downstream Task Datasets}
\textbf{EPIC-KITECHENS-100.} EPIC-KITECHENS-100 \cite{damen2020rescaling} is a large-scale egocentric audio-visual dataset, which captures daily activities in the kitchen. The videos are being recorded in 45 different kitchens and contain 100 hours of data. It includes 90K trimmed action clips and they capture the hand-object activities. The ground-truth labels are formed by a verb and a noun (e.g., move tap, open kettle, and open bin). There are 97 verb classes and 300 noun classes in total. Most of the actions are short-duration (average action length is 2.6 seconds). The audio is sampled at 24kHz.

\textbf{EPIC-Sound.} EPIC-Sound \cite{huh2023epic} is a large-scale audio event classification dataset that is re-annotated from the EPIC-KITCHENS-100 dataset. Unlike EPIC-KITCHENS-100, the actions in EPIC-Sound can be discriminated purely from the audio, for example, cutting food instead of cutting tomatoes. This dataset includes 75.9k segments of audible events and actions with 44 classes.

\textbf{Speech Commands V2 (KS2).} Speech Commands V2 \cite{speechcommandsv2} is an automatic speech commands recognition dataset that contains more than 100K audio clips with 1 second for each of them. It also contains 35 common speech commands for the recognition task. 

\textbf{UrbanSound8K (Urban8K).} UrbanSound8K \cite{Salamon:UrbanSound:ACMMM:14} is a urban sound event classification dataset. It contains 8K labeled sound excerpts with less than 4 seconds for each clip and 10 urban sound classes. 

\textbf{NSynth.} Nsynth \cite{engel2017neural} is an audio dataset containing 305,979 musical notes and each of them with a unique pitch, timbre, and envelope. The sounds were collected from 1006 instruments from commercial sample libraries and annotated based on their source, instrument family, and sonic qualities. There are 11 instrument families in total. 

\subsection{Training and Validation Details}
\label{subsec:trian-valid-details}
We follow the same training strategy as the baseline Slow-Fast model \cite{kazakos2021slow}, i.e., we pretrain our models on the VGG-SOUND dataset \cite{chen2020vggsound} followed by fine-tuning it on the downstream tasks datasets. The input audio signal is first converted into a Log-Mel spectrogram with 128 Mel bands before feeding it to the network. During both the pretraining and fine-tuning stages, we applied the augmentation methods proposed in SpecAugment \cite{park2019specaugment} following the common practice as in \cite{kazakos2021slow, gong2021ast, gong2022ssast, chen2022hts}. These augmentations include frequency masking, time masking, and time warping. All the models are trained with a batch size of 32 on 4x NVIDIA RTX3090 GPUs during the pretraining and fine-tuning stages. 

\subsubsection{Pretraining}
For the pretraining stage, we follow the baseline Slow-Fast model experiment setting \cite{kazakos2021slow} and randomly pick a sample of 5.12 seconds from the audio signal followed by feeding it to the Log-Mel filter banks with a window size of 20ms, and a hop length of 10ms. This results in a spectrogram of size 512 × 128, which is given as an input to the model. We follow the training setting described in \cite{kazakos2021slow} to train the models using an SGD optimizer for 50 epochs with a momentum of 0.9 and an initial learning rate of 0.01. We drop the learning rate by 0.1 at epochs 30 and 40.

\subsubsection{Fine-Tuning}
 We use the same strategy as \cite{kazakos2021slow} in the fine-tuning stage. We attach a linear prediction head on top of the VGG-Sound pre-trained backbone model to classify the target classes in different fine-tuning datasets. We froze all the batch normalization layers except the first one in the stem layer and fine-tuned the whole model. We use the same optimizer setting as the pretraining stage, except the initial learning rate is set to 0.001, which is reduced after 20 and 25 epochs by a factor of 0.1. The model is finetuned for 30 epochs. 

 \begin{table*}[t]
        \caption{Comparison with CNN-Based SOTA methods on VGG-Sound event classification pretraining dataset. Reparam stands for re-parameterization. Param stands for parameter size. GFLOPs stands for floating point operations. TP stands for throughput. Top-1 stands for top-1 accuracy, Top-5 stands for Top-5 accuracy, mAP stands for mean average precision and AUC stands for area under curve. Note that the models are separated by the model size. The upper three rows represent the small model size, while the remaining rows represent the large model size.}
        \centering
        \resizebox{2\columnwidth}{!}{
        \begin{tabular}{l|c|c|c|c|c|c|c|c|c|c}
            \hline
             \textbf{Model} & Reparam (M) & \textbf{Param (M)} & \textbf{GFLOPs} & \textbf{TP (GPU)} & \textbf{TP (CPU)} & \textbf{Top-1} & \textbf{Top-5} & \textbf{mAP} & \textbf{AUC} & \textbf{d-prime} \\
            \hline
            BCResNets-8 \cite{kim2021broadcasted} & N/A & 0.39 & 1.45 & 492 & 5.62 & 46.42 & 73.70 & 35.6 & 95.8 & 2.45 \\
            BN-InceptionNetV1 \cite{ioffe2015batch} & N/A & 6.35 & 1.86 & 2031 & 12.10 & \textbf{50.85} & \textbf{77.20} & \textbf{53.2} & 97.3 & 2.73 \\
            AudioRepInceptionNeXt-B0 (ours) & \xmark & 2.18 & 0.49 & 1399 & 11.80 & 49.25 & 76.33 & 52.1 & \textbf{97.5} & \textbf{2.77} \\
            \rowcolor{gray!20}
            AudioRepInceptionNeXt-B0 (ours) & \cmark &\textbf{2.11} & \textbf{0.46} & \textbf{2245} & \textbf{16.40} & 49.25 & 76.33 & 52.1 & \textbf{97.5} & \textbf{2.77} \\
            \hline
            Slow-Fast (baseline)\cite{kazakos2021slow} & N/A & 26.68 & 5.55 & 796 & 6.90 & \textbf{52.24} & \textbf{78.14} & \textbf{54.4} & 97.5 & 2.76 \\
            RepLKNet-31T \cite{ding2022scaling} & \xmark & 29.52 & 7.86 & 279 & 0.41 & 52.22 & 78.03 & 54.4 & 97.5 & 2.78\\
            RepLKNet-31T \cite{ding2022scaling} & \cmark & 29.32 & 7.79 & 295 & 0.43 & 52.22 & 78.03 & 54.4 & 97.5 & 2.78 \\
            ResNet50 \cite{jansen2018unsupervised} & N/A & 24.13 & 5.26 & 915 & \textbf{7.80} & 52.07 & 77.72 & 54.1 & 97.3 & 2.74 \\
            InceptionNeXt-Tiny \cite{yu2023inceptionnext} & N/A & 24.20 & 5.46 & 682 & 7.70 & 50.16 & 76.28 & 52.5 & 97.4 & 2.75  \\
            AudioRepInceptionNeXt-B1(2D) & \xmark & 13.70 & 3.52 & 529 & 0.91 & 51.26 & 77.54 & 53.4 & 97.5 & 2.77 \\
            AudioRepInceptionNeXt-B1 (2D) & \cmark & 13.19 & 3.27 & 666 & 0.95 & 51.26 & 77.54 & 53.4 & 97.5 & 2.77 \\
            AudioRepInceptionNeXt-B1 (ours) & \xmark & 11.83 & 2.62 & 700 & 6.10 & 51.96 & 77.86 & 54.0 & 97.6 & 2.79\\
            \rowcolor{gray!20}
            AudioRepInceptionNeXt-B1 (ours) & \cmark & \textbf{11.69} & \textbf{2.55} & \textbf{1019} & 7.50 & 51.96 & 77.86 & 54.0 & \textbf{97.6} & \textbf{2.79} \\
            \hline
        \end{tabular}
        }
    \label{tab:baseline-comparsion}
\end{table*}

For the EPIC-KITECHENS-100, we follow the setting described in \cite{kazakos2021slow} and randomly pick 2.08 seconds of audio and apply a Log-Mel-Filter bank with a window size of 10ms and a hop length of 5ms. This results in a spectrogram of size 416×128 which is fed to the model as an input. Note that due to the need to downsample the spectrogram by a factor of 32 in our proposed model, adjustments were made to the sampling time, resulting in a slightly longer duration of 2.08 seconds. For EPIC-SOUND, we follow the setting described in \cite{huh2023epic} and apply a mel-log-filter bank with a window size of 10ms and a hop length of 5ms. However, we randomly pick 2.08 seconds instead of 2 seconds as stated in \cite{huh2023epic} to account for the downsampling stage in our model. This results in a spectrogram of size 416×128. For KS2 dataset, which has a maximum audio length of 1.023 seconds, we use a similar windows size of 5ms and hop length of 2ms in \cite{huh2023epic}. This results in a spectrogram of size 512×128.  For NSYNTH and Urban8K datasets, following the setup for the VGG sound dataset in \cite{kazakos2021slow}, we randomly pick a sample of 4.16 seconds from the audio signal and apply a Log-Mel-Filter bank with a window size of 20ms and a hop length of 10ms. This results in a spectrogram of size 416×128.

\subsection{Evaluation Metrics}
For the evaluation of the VGG-Sound classification task, we follow the protocol of \cite{chen2020vggsound, hershey2017cnn, kazakos2021slow} and report the top-1 accuracy, top-5 accuracy, mean average precision (mAP), area under curve (AUC) and d-prime. For the evaluation of the EPIC-Sounds, KS2, Urban8K, and Nsynth, we report the top-1 and top-5 accuracy. For the EPIC-KITCHENS-100, we follow the evaluation method in \cite{damen2020rescaling} and report the top-1 and top-5 accuracy of verb and noun classes. Additionally, we report the top-1 and top-5 accuracy on unseen audio clips in EPIC-KITCHENS-100 to test the generalization ability of the fine-tuned model. 

For the measurement of the GPU inference speed (sample/secs), we test all the models on an NVIDIA RTX3090 using a batch size of 32. We first feed 50 batches to warm up the hardware, followed by 50 batches to record the average running time. To measure the CPU inference speed, we conduct the tests on Intel\textregistered Xeon\textregistered Gold 6226R CPU @ 2.90GHz using a batch size of 1. The tests are performed using a single thread, and we record the average time after the 50 rounds.

\subsection{Effects of Re-parametrization}
\label{subsec:effect-of-reparam}
To verify the effectiveness of the re-parametrization in our proposed AudioRepInceptionNeXt, we perform the comparison against the state-of-the-art (SOTA) methods. The comparison is performed in terms of the parameter size, GFLOPs, throughput, and top-1 accuracy before and after the re-parametrization. We report the results in Table \ref{tab:baseline-comparsion}. One can see that before reparameterization, AudioRepInceptionNeXt-B1 is 12\% and 23\% slower than the Slow-Fast \cite{kazakos2021slow} and ResNet50 \cite{jansen2018unsupervised} respectively, and 60\% faster than RepLKNet \cite{ding2022scaling}. We conjecture the low throughput of AudioRepInceptionNeXt is due to its complicated multi-branch design although it incurs lower parameters and theoretical GFLOPs than SlowFast and ResNet50. As discussed in \cite{ma2018shufflenet, ding2021repvgg, ding2022scaling}, the large kernel with depthwise convolution increases the memory access costs. Additionally, as mentioned in \cite{ma2018shufflenet}, small operators (i.e., individual convolution (e.g., $1\times 1$) and pooling operations) with multi-branch design are less efficient on GPUs and introduce extra kernel launching and synchronization overhead thus reduces the degree of parallelism on GPU. In our AudioRepInceptionNeXt block, the separable kernels of size $1 \times 11$, $1 \times 3$, $11 \times 1$, and $3 \times 1$  are relatively small operators compared to $1 \times 21$  and $21 \times 1$. To address these issues, we employ the re-parametrization technique, as discussed in Section \ref{subsec:rep-inference-model}, that results in a network style similar to ResNet50, as shown in Figure \ref{fig:AudioConvNeXt-block-b}. One can see from Table \ref{tab:baseline-comparsion} that the reparametrized version of AudioRepInectionNeXt achieves $1.28\times$, $1.11\times$, and $3.65\times$ improvement in throughput compared to the Slow-Fast, ResNet50, and RepLKNet respectively while maintaining comparable performance. Notably, the re-parametrization technique does not impact the accuracy, and results in lossless compression of the model. In the rest of the sections, we report the evaluation of the reparametrized version of AudioRepInceptionNeXt.

\subsection{Comparison against the CNN-based baselines on VGG Sound pretraining}
\label{sec:performance-on-pretrain} We compare the performance of AudioRepInceptionNeXt against the CNN-Based baselines that include BCResNet-8 \cite{kim2021broadcasted}, BN-InceptionNet \cite{ioffe2015batch}, ResNet \cite{jansen2018unsupervised}, Slow-Fast model \cite{kazakos2021slow}, InceptionNeXt \cite{yu2023inceptionnext}, RepLKNet \cite{ding2022scaling} and 2D AudioRepInceptionNeXt(without any separable convolution kernel and with branch configurations of $21 \times 21$, $11 \times 11$ and $3 \times 3$ kernels) as shown in Figure \ref{fig:AudioConvNeXt-block-b} (Right). To ensure a fair comparison, we downscaled the original RepLKNet-31B model, with 79M parameters, to the RepLKNet-31T model with 29.3M parameters. This downsizing involved reducing the channel size to 64, 128, 320, and 512 for model stages 1 to 4, respectively. All the evaluation is performed on the VGG-Sound event classification dataset. We report the results in Table \ref{tab:baseline-comparsion}. 

\begin{table*}[t]
        \caption{Transfer learning results on EPIC-Sounds, KS2, Urban8K and Nsynth. Note that the models are separated by the model size. The upper three rows represent the small model size, while the remaining rows represent the large model size.}
        \centering
        \resizebox{2\columnwidth}{!}{
        \begin{tabular}{l|cc|cc|cc|cc|cc}
            \hline
            \multirow{2}{*}{Model} & \multirow{2}{*}{Param (M)} & \multirow{2}{*}{GFLOPs} & \multicolumn{2}{c|}{EPIC-Sounds} & \multicolumn{2}{c|}{KS2}  & \multicolumn{2}{c|}{Urban8K} & \multicolumn{2}{c}{Nsynth} \\
             &  &  & Top-1 & Top-5 & Top-1 & Top-5 & Top-1 & Top-5 & Top-1 & Top-5\\
            \hline
            BCResNet-8 \cite{kim2021broadcasted} & 0.33 & 1.17 & 52.91 & 84.08 & 95.80 & 99.19 & \textbf{84.82} & \textbf{98.73} & 76.13 & 95.44 \\
            BN-InceptionNetV1 \cite{ioffe2015batch} & 6.07  & 1.51 & \textbf{53.81} & \textbf{85.00} & 96.84 & 99.30 & 82.41 & 98.39 & 78.09 & 94.21\\
            \rowcolor{gray!20}
            AudioRepInceptionNeXt-B0 & \textbf{2.02} & \textbf{0.37} & 53.43 & 84.77 & \textbf{97.11} & \textbf{99.45} & 82.33 & 98.15 & \textbf{78.94} & \textbf{96.58} \\
            \hline
            Slow-Fast (baseline) \cite{kazakos2021slow} & 26.06  & 4.50  & \textbf{52.84} & 83.12 & \textbf{97.30} & 99.37 & 81.83 & 97.13 & 77.49 & 96.32  \\
            RepLKNet-31T \cite{ding2022scaling} & 29.18  &  6.33 & 52.79 & 82.40 & 97.01  & \textbf{99.48} & 83.40 & 98.10 & \textbf{78.48} & 96.75 \\
            ResNet50 \cite{jansen2018unsupervised} & 23.59 & 4.27  &  52.57 & 82.77 & 97.17 & 99.37 & 80.62 & 97.96 & 77.56 & 96.36\\
            InceptionNeXt-Tiny \cite{yu2023inceptionnext} & 24.00  & 4.43  & 51.24 & 81.73 &  96.94 & 99.34 & 82.08 & 97.54 & 77.24 & 96.68 \\
            AudioRepInceptionNeXt-B1 (2D) & 13.05 & 2.65 & 51.97 & 82.35 & 96.83 & 99.38 & 81.93 & 97.75 & 77.33 & \textbf{97.47} \\
            \rowcolor{gray!20}
            AudioRepInceptionNeXt-B1 (ours) & \textbf{11.55} & \textbf{2.07} & 52.74 & \textbf{83.22} & 97.20 & 99.39 & \textbf{83.44} & \textbf{98.13} & 77.28 & 97.00 \\
            \hline
        \end{tabular}
        }
    \label{tab:transfer-ks2-urban}
\end{table*}

\begin{table*}[t]
    \caption{Transfer learning results on EPIC-KITCHENS-100.Note that the models are separated by the model size. The upper three rows represent the small model size, while the remaining rows represent the large model size.}
    \centering
    \resizebox{2\columnwidth}{!}{
    \begin{tabular}{l|cc|cccccc|ccc}
        % \toprule
        \hline
        & & & \multicolumn{6}{c|}{Overall} & \multicolumn{3}{c}{Unseen Participants} \\
        &  &  & \multicolumn{3}{c}{Top-1 Accuracy} & \multicolumn{3}{c|}{Top-5 Accuracy} & \multicolumn{3}{c}{Top-1 Accuracy}\\
        Model & Param (M) & GFLOPs & Verb & Noun & Action & Verb & Noun & Action & Verb & Noun & Action \\
        \hline
        BCResNet-8 \cite{kim2021broadcasted} & 0.41 & 1.17 & 42.26 & 18.14 & 1.99 & 77.72 & 42.00 & 2.19 & 37.18 & 17.84 & 2.44 \\
        BN-inceptionNetV1 \cite{ioffe2015batch} & 6.44 & 1.51 & 44.68 & \textbf{21.03} & 12.90 & 79.94 & 46.36 & 27.77 & 38.30 & 17.27 & 9.67 \\
        \rowcolor{gray!20}
        AudioRepInceptionNeXt-B0 & \textbf{2.14} & \textbf{0.37} & \textbf{46.34} & 20.71 & \textbf{13.51} & \textbf{80.52} & \textbf{46.75} & \textbf{29.77} & \textbf{40.85} & \textbf{17.28} & \textbf{9.86} \\
        \hline
        Slow-Fast (baseline) & 26.88 & 4.33 & 46.86 & \textbf{22.98} & 15.52 & 80.12 & 47.58 & 30.17 & 39.62 & 17.28 & \textbf{10.52} \\
        RepLKNet-31T \cite{ding2022scaling} & 29.36 & 6.33 & 47.33 & 23.46 & \textbf{16.12} & 80.25 & 47.95 & 30.75 & 40.00 & \textbf{17.37} & 9.48 \\
        ResNet50 \cite{jansen2018unsupervised} & 24.32 & 4.28 & 46.03 & 22.79 & 15.21 & \textbf{80.71} & 47.83 & 30.06 & \textbf{40.84} & 16.24 & 9.76 \\
        InceptionNeXt-Tiny \cite{yu2023inceptionnext} & 24.27 & 4.43 & 44.72 & 21.80 & 14.07 & 79.27 & 45.93 & 28.18 & 38.77 & 15.96 & 9.10 \\
        AudioRepInceptionNeXt-B1 (2D) & 13.23 & 2.65 & 46.89 & 22.37 & 15.01 & 79.95 & \textbf{48.12} & 31.09 & 40.46 & 16.99 & 9.57  \\
        \rowcolor{gray!20}
        AudioRepInceptionNeXt-B1 & \textbf{11.73} & \textbf{2.07} & \textbf{47.57} & 22.14 & 15.42 & 80.45 & 48.06 & \textbf{31.17} & 39.62 & 17.00 & 9.57 \\
        \hline
    \end{tabular}
    }
    \label{tab:transfer-epic-kitchens}
\end{table*}

Our key findings include: First, the proposed AudioRepInceptionNeXt demonstrates a remarkable reduction in the number of parameters and theoretical GFLOPs while achieving comparable or superior performance compared to other CNN-based methods. For instance, when compared to the multi-stream Slow-Fast model, AudioRepInceptionNeXt-B1 achieves a 56\% reduction in parameters and 54\% reduction in theoretical GLOPs, with only a slight performance drop of 0.28\%. Similarly, in comparison to multi-branch large kernel InceptionNeXt-Tiny, our proposed model achieves a 52\% reduction in parameters and a 53\% reduction in GFLOPs while attaining a higher accuracy of 1.8\%. Furthermore, when compared to 2D large kernel-based RepLKNet-31T, our model achieves comparable accuracy with a difference of only 0.26\% while saving 71.05\% of parameters and 68\% of GFLOPs. Additionally, compared to the 2D version of the proposed AudioRepInceptionNeXt-B1, our model achieves a 0.7\% higher accuracy while saving 11\% of model parameters and 22\% of GFLOPs. This result is consistent with the findings in \cite{kazakos2021slow, xiao2020audiovisual}, demonstrating that the model with separable kernels performs better by enabling the extraction of temporal and frequency information independently. Second, our model demonstrates the fastest inference speed on GPU compared to other CNN-based baselines, aligning with the theoretical GFLOPs. The proposed AudioRepInceptionNeXt-B0 obtains 2245 fps (vs. 2031 fps and 492 fps obtained by BN-InceptionNetV1 and BCResNet-8, respectively), while AudioRepIncetpionNeXt-B1 obtains 1019 fps (vs.796 fps obtained by Slow-Fast). Third, the proposed AudioRepInceptionNeXt-B1 achieves a higher inference speed on the CPU compared to the Slow-Fast model, RepLKNet-31T, AudioRepInceptionNeXt-B1 2D, and BN-InceptionNetV1, while maintaining a comparable performance with Slow-Fast and RepLKNet-31T. We also note that the CPU inference speed of AudioRepInceptionNeXt-B1 is comparable to ResNet50 and InceptionNeXt-Tiny while surpassing InceptionNeXt-Tiny in terms of accuracy and achieving competitive accuracy compared to ResNet50. When comparing the inference speeds of ResNet50 and AudioRepInceptionNeXt-B1 on CPU and GPU, 
we conjecture that the difference in the inference speed among these models can be attributed to the optimized implementation of depthwise convolution on GPU compared to CPU, as discussed in \cite{lu2021optimizing}. Additionally, the disparity in arithmetic intensity (the ratio of compute to memory operations), as mentioned in Section \ref{subsec:effect-of-reparam}, also contributes to the observed variations in speed on GPU and CPU. These findings highlight the superior performance of AudioRepInceptionNeXt in terms of parameter efficiency, theoretical GFLOPs, and inference speed compared to other CNN-based baselines.  

\begin{table*}[t]
        \caption{Model size and Inference time comparison between CNN-Based methods and AudioRepInceptionNeXt on the mobile device. Lower inference time is better. Note that the models are separated by the model size. The upper three rows represent the small model size, while the remaining rows represent the large model size.}
        \centering
        \resizebox{1.5\columnwidth}{!}{
        \begin{tabular}{l|c|c|c|c}
            \hline
             Model & Param (M) & GFLOPs & Model Size (MB) & Inference Time (ms)\\
            \hline
            BCResNet-8 \cite{kim2021broadcasted} & \textbf{0.33} & 1.17 & \textbf{1.6} & 227 \\
            BN-InceptionNetV1 \cite{ioffe2015batch} & 6.35 & 1.86 & 25 & 114 \\
            \rowcolor{gray!20}
            AudioRepInceptionNeXt-B0 (ours) & 2.11& \textbf{0.46} & 9 & \textbf{59}\\
            \hline
            Slow-Fast (baseline) \cite{kazakos2021slow} & 26.68 & 5.55 & 106 & 317 \\
            RepLKNet-31T \cite{ding2022scaling} & 29.32 & 7.79 & 117 & 2136\\
            ResNet50 \cite{jansen2018unsupervised} & 24.13 & 5.26 & 96 & 310 \\
            InceptionNeXt-Tiny \cite{yu2023inceptionnext} & 24.20 & 5.46 & 97 & 357 \\
            AudioRepInceptionNeXt-B1 (2D) & 13.19 & 3.27 & 53 & 615 \\
            \rowcolor{gray!20}
            AudioRepInceptionNeXt-B1 (ours) & \textbf{11.69} & \textbf{2.55} & \textbf{47} & \textbf{232} \\
            \hline
        \end{tabular}
        }
    \label{tab:mobile-runtime-comparsion}
\end{table*}

\begin{figure}[t]
\centering
\includegraphics[width=0.9\linewidth]{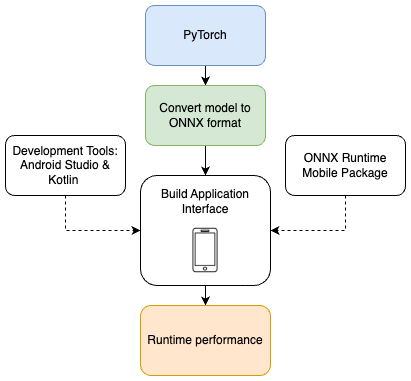}
\caption{Mobile runtime application development flow by using the ONNX Runtime library.}
\label{fig:mobile-eva-step}
\end{figure}

\subsection{Performance on Downstream task datasets}
To verify the conclusion from Section \ref{sec:performance-on-pretrain}, we conduct transfer learning experiments on multiple datasets: Speech Command V2 \cite{speechcommandsv2}, UrbanSound8K \cite{salamon2014dataset}, EPIC-KITCHEN-100\cite{damen2020rescaling}, NSynth \cite{engel2017neural} and EPIC-Sound \cite{huh2023epic}. As shown in Table \ref{tab:transfer-ks2-urban} and Table \ref{tab:transfer-epic-kitchens}, our proposed model achieves comparable performance to the multi-stream Slow-Fast model in terms of accuracy on four transfer learning datasets, while saving 56\% and 54\% of parameters and GFLOPs, respectively. Moreover, when compared to the Urban8K dataset our model outperforms the Slow-Fast model by 1.61\% in terms of top-1 accuracy. Additionally, when compared to other CNN-based methods, our methods achieve comparable or superior performance in terms of accuracy, while having the lowest parameters count and computational GFLOPs. These results indicate that our model can learn the rich representations that are applicable to different domains and are robust when transferred to various audio understanding tasks. These findings are also aligned with the results obtained from the pretraining on the VGG Sound dataset, as discussed in section \ref{sec:performance-on-pretrain}.

\subsection{Implementation on Mobile Devices}
To evaluate the inference speed on the mobile device, we deploy all models on an Android mobile platform. The implementation procedure is summarized in Fig.\ref{fig:mobile-eva-step}. We first convert the PyTorch models to the Open Neural Network Exchange (ONNX) format \cite{bai2019onnx}. It is an open-source machine-independent format compatible with different hardware, drivers, and operating systems. We then utilize the Android development tools, specifically Android Studio and Kotlin, to build an application interface for conducting the speed evaluation. To run the ONNX models on Android mobile devices, we leverage the ONNX Runtime mobile package. It provides a lightweight and optimized runtime specifically designed for mobile platforms. During the evaluation, we run all models on Redmi Note 9 Pro smartphone with Snapdragon 720G SoC with floating point 32. Specifically, we feed an input with a batch size of 1 and record the average inference time by processing 50 batches.

\subsection{Runtime Performance on Mobile Devices}
As shown in Table \ref{tab:mobile-runtime-comparsion}, our model demonstrates superior performance in terms of inference speed and model size on mobile devices compared to other state-of-the-art CNNs. Notably, the AudioRepInceptionNeXt-B1 model outperforms the classical ResNet50 by reducing the inference time and model size by 25\% and 50\%, respectively. In comparison to the recent multi-branch large kernel InceptionNeXt-Tiny, our model achieves a 35\% faster inference speed and 51\% smaller model size. When compared to RepLKNet-31T, a 2D large kernel-based model, our proposed model exhibits substantial improvements, saving 89\% of the inference time and reducing the model size by 60\%. Moreover, our method surpasses the multi-stream Slow-Fast model, achieving 27\% lower inference time and 56\% smaller model size. The results clearly demonstrate the efficiency of our model in terms of both inference speed and model size on mobile devices, showcasing its superiority over existing state-of-the-art CNNs.

\begin{table*}[t]
        \caption{Ablation study on the design of multi-branch large kernel.}
        \centering
        \resizebox{2\columnwidth}{!}{
        \begin{tabular}{l|c|c|c|c|c|c|cc|cc|cc|c|c|c|c|c}
            \hline
             \multirow{2}{*}{Structure} & \multirow{2}{*}{$3 \times 3$}  & \multirow{2}{*}{$11 \times 11$} & \multirow{2}{*}{$21 \times 21$} & \multirow{2}{*}{$31 \times 31$} & \multirow{2}{*}{Identity} & \multirow{2}{*}{Inverted Bottleneck} & \multicolumn{2}{c|}{Param (M)} & \multicolumn{2}{c|}{GFLOPs}  & \multicolumn{2}{c|}{Throughput} & \multirow{2}{*}{Top-1} & \multirow{2}{*}{Top-5} & \multirow{2}{*}{mAP} & \multirow{2}{*}{AUC} & \multirow{2}{*}{d-prime} \\
             & & & & & & & Before Rep. & After Rep. & Before Rep. & After Rep. & Before Rep. & After Rep. &  &  &  &  & \\
            \hline
            s1  & \cmark & \xmark & \xmark & \xmark & \cmark & \cmark & 11.56 & 11.54 & 2.49 & 2.48 & 1118 & 1175 & 51.40 & 77.79 & 53.82 & 97.43 & 2.75 \\
            s2  & \xmark & \cmark & \xmark & \xmark & \cmark & \cmark & 11.62 & 11.60 & 2.52 & 2.51 & 1048 & 1101 & 51.55 & 77.46 & 53.76 & 97.50 & 2.77 \\
            s3  & \xmark & \xmark & \cmark & \xmark & \cmark & \cmark &  11.69 & 11.68 & 2.55 & 2.55 & 971 & 1019 &  51.36 & 77.83 & 53.54 & 97.61 & 2.79 \\
            s4  & \cmark & \xmark & \cmark & \xmark & \cmark & \cmark & 11.73 & 11.69 & 2.57 & 2.55 & 833 & 1019 & 51.60 & 77.74 & 53.87 & 97.55 & 2.78 \\
            s5 & \xmark & \cmark & \cmark & \xmark & \cmark & \cmark & 11.79  & 11.69 &  2.60 & 2.55  & 794 & 1019 & 51.64 & 77.68 & 53.66 & 97.55 & 2.78 \\
            s6 & \cmark & \cmark & \cmark & \xmark & \cmark & \cmark & 11.83  & 11.69 & 2.62 & 2.55 & 701 & 1019 & \textbf{51.96} & \textbf{77.86} & 53.97 & 97.58 & 2.79 \\
            s7 & \cmark & \cmark & \cmark & \cmark & \cmark & \cmark & 12.08  & 11.69 & 2.74 & 2.55 & 554 & 1001 & 51.49 & 77.77 & \textbf{54.23} & 97.58 & 2.79 \\
            s8 & \cmark & \cmark & \cmark & \xmark & \xmark & \cmark & 11.83  & 11.69 & 2.62 & 2.55 & 727 & 1076 & 51.29 & 77.44 & 53.68 & \textbf{97.62} & \textbf{2.80} \\
            s9 & \cmark & \cmark & \cmark & \xmark & \cmark & \xmark & 3.02  & 2.85 & 0.73 & 0.65 &  1011 & 1803 & 51.02 & 77.41 & 53.67 & 97.61 & 2.79 \\

            \hline
        \end{tabular}
        }
    \label{tab:ablation-studies-multi-branch}
\end{table*}

\section{Ablation Studies}
\label{sec:Ablation Studies}

We conduct a series of ablation studies on AudioRepInceptionNeXt to verify the significance of multi-branch design and the usage of large kernels. To save the compute, we only conduct ablation studies on the VGG-Sound dataset and AudioRepInceptionNeXt-B1 architecture. Specifically, we first ablate some branches with different kernel sizes and then observe the performance changes. We then compare the AudioRepInceptionNeXt block to a counterpart with a block without a shortcut path. Concretely, we remove the identity shortcut for both horizontal and vertical multi-scale kernels. Note that we follow the same training setting as mentioned in section \ref{sec:experiment}. We report the results of such an ablation study in terms of the number of parameters, GFLOPs, throughput, and accuracy before and after re-parameterization in Table \ref{tab:ablation-studies-multi-branch}. 

As shown in Table \ref{tab:ablation-studies-multi-branch}, for the single branch setting (i.e., structure s1 to s3), there is no significant accuracy improvement when we enlarge the kernel size from 3 to 11. Meanwhile, there is 0.19\% performance degradation when we further enlarge the kernel size from 11 to 21. We conjecture that the model overlooks the local details when we enlarge the respective field of the kernel. To verify it, we introduce the multi-branch setting (i.e., structure s4 to s6) with kernel sizes 11 and 3 in our ablation studies. The results demonstrate that all the multi-branch models can lift the accuracy above 51.55\% and outperform the single-branch models. Comparing the triple branch (i.e., structure s6) to the dual branch (i.e., structure s4 and s5), we note that removing any single branch degrades the performance, suggesting that all the branches are indispensable. However, when we introduced additional branches with a kernel size of $31 \times 31$, the accuracy started to degrade by 0.4\%.

To verify the importance of the identity shortcut in the parallel multi-scale horizontal and vertical kernel, we remove all the identity shortcut layers and form a new structure s8 as shown in table \ref{tab:ablation-studies-multi-branch}. Compared to the full structure s6, the accuracy of s8 is dropped by 0.67\%. This indicates the importance of identity shortcuts in the AudioRepInceptionNeXt block.

To further verify the importance of the inverted bottleneck layer, we remove the $1 \times 1$ channel expansion layer and form a new structure s9 as shown in table \ref{tab:ablation-studies-multi-branch}. Compared to the full structure s6, the accuracy of s9 is dropped by 0.94\%. This demonstrates the importance of the inverted bottleneck.

\section{Conclusion}
\label{sec:conclusion}
In this study, we address the issue of computational and memory inefficiency in the multi-stream and multi-branch convolutional neural networks (CNNs). To alleviate these problems, a simple single-stream model was proposed which employs a parallel multi-scale separable kernel design, effectively reducing the number of parameter counts and GFLOPs by 50\% during the training. In order to eliminate the side effects arising from multi-scale kernel design, such as slow inference speed and frequent memory access, we utilize a reparameterization technique during the inference. Moreover, we adopt depthwise separable kernels and an inverted bottleneck design to address the inefficiencies associated with the large-scale kernels in the parallel branches of the multi-scale kernel. Our experimental results show that the proposed AudioRepInceptionNeXt achieves a favorable trade-off between parameter size and computational time while maintaining comparable or superior performance in relation to the baseline Slow-Fast model and AudioRepInceptionNeXt (2D). Additionally, our findings demonstrate that the proposed model is a robust learner capable of achieving better or comparable performance compared to the SlowFast model across various transfer learning datasets. This study provides valuable insights for researchers and practitioners in the field of deep learning who seek to enhance the computational efficiency of audio classification models.

%  % argument is your BibTeX string definitions and bibliography database(s)
% %\bibliography{IEEEabrv,../bib/paper}
% %
\bibliographystyle{IEEEtran}
\bibliography{egbib}

\vfill

\end{document}